# Rules and mechanisms governing octahedral tilts in perovskites under pressure

H. J. Xiang[1,2,3*], Mael Guennou[4], Jorge Íñiguez[4], Jens Kreisel[4,5], L. Bellaiche[2*]

[1]*Key Laboratory of Computational Physical Sciences (Ministry of Education), State Key Laboratory of Surface Physics, and Department of Physics, Fudan University, Shanghai 200433, P. R. China*

[2]*Physics Department and Institute for Nanoscience and Engineering, University of Arkansas, Fayetteville, Arkansas 72701, USA*

[3]*Collaborative Innovation Center of Advanced Microstructures, Nanjing 210093, P. R. China*

[4]*Materials Research and Technology Department, Luxembourg Institute of Science and Technology (LIST), 41 Rue du Brill, L-4422 Belvaux, Luxembourg*

[5]*Physics and Materials Science Research Unit, University of Luxembourg, 41 Rue du Brill, L-4422 Belvaux, Luxembourg*

Subject Areas: Computational Physics, Condensed Matter Physics, Materials Science

*email: hxiang@fudan.edu.cn, laurent@uark.edu

**ABSTRACT**

The rotation of octahedra (octahedral tilting) is common in $ABO_3$ perovskites and relevant to many physical phenomena, ranging from electronic and magnetic properties, metal-insulator transitions to improper ferroelectricity. Hydrostatic pressure is an efficient way to tune and control octahedral tiltings. However, the pressure behavior of such tiltings can dramatically differ from one material to another, with the origins of such differences remaining controversial. In this work, we discover several new mechanisms and formulate a set of simple rules that allow to understand how pressure affects oxygen octahedral tiltings, via the use and analysis of first-principles results for a variety of compounds. Besides the known A-O interactions, we reveal that the interactions between specific B-ions and oxygen ions contribute to the tilting instability. We explain the previously reported trend that the derivative of the oxygen octahedral tilting with respect to pressure ($dR/dP$) usually decreases with both the tolerance factor and the ionization state of the A-ion, by illustrating the key role of A-O interactions and their change under pressure. Furthermore, three new mechanisms/rules are discovered, namely that (i) the octahedral rotations in $ABO_3$ perovskites with empty low-lying d states on the B-site are greatly enhanced by pressure, in order to lower the electronic kinetic energy; (ii) $dR/dP$ is enhanced when the system possesses weak tilt instabilities, and (iii) for the most common phase exhibited by perovskites - the orthorhombic Pbnm state - the in-phase and antiphase octahedral rotations are not automatically both suppressed or both enhanced by the application of pressure, because of a trilinear coupling between these two rotation types and an antipolar mode involving the A-ions. We further predict that the polarization associated with the so-called hybrid improper ferroelectricity could be manipulated by hydrostatic pressure, by indirectly controlling the amplitude of octahedral rotations.

## I. INTRODUCTION

The perovskite structure is one of the most commonly occurring and important structural types in materials science. From both theoretical and applied points of view, perovskite materials are interesting since they display many diverse and intriguing properties, including superconductivity [1], colossal magnetoresistance [2], ferroelectricity [3], multiferroicity [4-6], or photovoltaicity [7]. The ideal perovskite oxide $ABO_3$ structure adopts the cubic space group $Pm\bar{3}m$, with the A cation surrounded by twelve oxygen anions in a dodecahedral environment and the B cation octahedrally coordinated with six oxygen ions. The perovskite structure can be viewed as a three-dimensional cubic network of corner-sharing $BO_6$ octahedra with the A cation sitting in the center of a cube defined by eight corner-sharing octahedral units. Although the ideal perovskite structure is cubic, most perovskite oxides are in fact distorted [8]. The most common type of distortion is octahedral rotation, i.e. rigid $BO_6$ octahedra tilts while maintaining their corner-sharing connectivity [9]. The octahedral rotation, which was believed to be due to the tendency to maximize the number of short A-O interactions [10], can have important effects on physical properties of perovskite compounds, particularly electrical and magnetic [11].

The effect of hydrostatic pressure on properties of perovskites has been investigated for a long time in condensed matter physics, solid state chemistry, materials science and earth science. For example, an external pressure causes the polar distortion of multiferroic TbMnO$_3$ to flop, and leads to the largest polarization values ever reported among spin-driven ferroelectrics [12]. It was also reported that hydrostatic pressure can significantly influence octahedral tilt angles. Regarding the pressure effect on octahedral tiltings, Samara *et al.* [13] proposed a rule in terms of the competition between the short-range Pauli repulsion and long-range Coulomb interactions in 1975. According to this picture, in the case of zone-boundary distortions (e.g., octahedral tiltings) the short-range interactions would increase with pressure much more rapidly than the long-range couplings, which should result in an increase of octahedral tiltings under pressure. This rule is in agreement with the pressure behaviors in orthorhombic CaSnO$_3$ [14] and CaTiO$_3$ [15], and tetragonal SrTiO$_3$ [16], However, this "general rule" is violated by experimental results of other materials: For example, rhombohedral LaAlO$_3$ [17], as well as orthorhombic YAlO$_3$, GdFeO$_3$, GdAlO$_3$ [18,19] and SmFeO$_3$ [20], all become less distorted under pressure. Note that the behavior in LaAlO$_3$ was confirmed in a first-principles study [21]. Later on, another empirical rule [22] based on the relative compressibility of the AO$_{12}$ and BO$_6$ polyhedra was proposed to account for the observed differences in behavior among various compounds. This rule states that, for perovskites in which the A cation has a lower formal charge than the B cation (e.g., MgSiO$_3$, CaSnO$_3$ or CaSiO$_3$), the AO$_{12}$ polyhedra are more compressible than the BO$_6$ octahedra and, as a result, the tilts of the BO$_6$ octahedra increase with pressure, thereby reducing the unit-cell volume. In contrast, whenever the A and B cations have the same formal charge (e.g., LaAlO$_3$ and GdFeO$_3$), the BO$_6$ octahedra are more compressible than the AO$_{12}$ polyhedra and, as a consequence, the tilts of the octahedra decrease with increasing pressure – thus evolving towards the cubic phase. However, the rule of Angel *et al.* is in conflict with (i) a density functional study predicting that pressure gradually reduces (rather than enhances) the tilting of the SiO$_6$ octahedra in orthorhombic CaSiO$_3$ [23]; (ii) first-principles calculations showing that the instability of antiphase tiltings becomes stronger (rather than weaker) with increasing pressure for the cubic phase of REAlO$_3$ compounds with small rare-earth (RE) ion (e.g., Er) [21]; and (iii) a recent Raman scattering and synchrotron powder X-ray diffraction study suggesting that the octahedral tilts may increase with pressure in rare-earth chromites RECrO$_3$ with small RE ions [24]. Note that the possible failure of this rule was also pointed out by Zhao *et al.* [25]. Therefore, the origin of the distinct pressure behaviors of octahedral tilting in perovskites remains puzzling.

In this work we aim at revealing and understanding the origin of the diverse pressure behaviors of octahedral tilting in perovskites, by conducting and analyzing first-principles calculations on many different and representative materials. We also report the discovery of new rules/effects pertaining to the effect of pressure on octahedral tilting. Not only does our work provide a unified set of rules on the effects of pressure, but also suggests original ways to tune these tiltings and, therefore, the properties of perovskites.

## II. RESULTS

As indicated in the Appendix, we perform density functional theory (DFT) simulations on a variety of perovskites under hydrostatic pressure.

### A. General trends for in-phase and antiphase tiltings

In order to understand the various effects that pressure can have on octahedral tiltings, we first focus on cases for which there is only a single type of tilting about a single pseudo-cubic axis. In other words, we consider two possibilities: an in-phase tilt about the pseudo-cubic [001] direction (i.e., $a^0a^0c^+$ in Glazer's notation[9]) and an antiphase tilt about the same axis (i.e., $a^0a^0c^-$). Figure 1 reports our results for the derivative of in-phase and antiphase tilting amplitude with respect to pressure (that is, $dR_{in}/dP$ and $dR_{anti}/dP$, respectively) as a function of the tolerance factor, for many different perovskite materials. Note that the tolerance factor [26] is defined as $t = \dfrac{r_A + r_O}{\sqrt{2}(r_B + r_O)}$ (where $r_A, r_B, r_C$ denote the radii [27] of the A-cation, B-cation, and

O-anion, respectively) and is widely used to discuss the stability of perovskite structures. Here, we consider (i) the REAlO$_3$, REFeO$_3$ and RECrO$_3$ families as representative of $A^{3+}B^{3+}O_3^{2-}$ materials; (ii) the CaBO$_3$ (with B = Ti, Zr, Hf, Si, Ge, Sn, Mn) and SrBO$_3$ (B = Ti, Ge, Mn, Zr, Hf) families as examples of $A^{2+}B^{4+}O_3^{2-}$ compounds; and (iii) LiNbO$_3$, LiTaO$_3$ and NaNbO$_3$ for $A^{1+}B^{5+}O_3^{2-}$ systems (we include LiNbO$_3$ and LiTaO$_3$ here merely for the comparison with NaNbO$_3$). Note that these materials are considered in idealized $a^0a^0c^+$ and $a^0a^0c^-$ structures so that we can investigate general trends in the pressure behavior of an individual tilting pattern, although most of them present more complicated tilting structures in reality.

As shown in Figs. 1a and 1b, the pressure behavior of in-phase tilts is rather similar to that of the antiphase tilts in all considered compounds. We will thus mainly discuss the antiphase case, as this pattern is the most common one among perovskites.

We find that, usually, $dR_{anti}/dP$ decreases with the tolerance factor for each series. For example, for REFeO$_3$, $dR_{anti}/dP$ decreases from $-2.0\times10^{-5}$ to $-4.3\times10^{-5}$ kbar$^{-1}$ [28] where RE varies from Lu to La. Another interesting trend is that $dR_{anti}/dP$ for the $A^{2+}B^{4+}O_3^{2-}$ and $A^{1+}B^{5+}O_3^{2-}$ families is larger than that for the $A^{3+}B^{3+}O_3^{2-}$ materials. In fact, $dR_{anti}/dP$ is negative for all REAlO$_3$ and REFeO$_3$ compounds, i.e., pressure suppresses the antiphase octahedral tilting in these cases. In contrast, $dR_{anti}/dP$ is positive for the $A^{1+}B^{5+}O_3^{2-}$ compounds and for most of the $A^{2+}B^{4+}O_3^{2-}$ family members. Note also that CaSiO$_3$, SrGeO$_3$ and SrMnO$_3$ have negative $dR_{anti}/dP$ and relatively large tolerance factors, which is at odds with the rule proposed by Angel *et al.* [22].

Figure 1 thus indicates that the pressure behavior of the tilting angle can be typically understood if (1) one considers the decrease of $dR_{anti}/dP$ with the tolerance factor within a family series, including the possibility that $dR_{anti}/dP$ changes sign as the tolerance factor increases; and (2) for similar values of the tolerance factor, $A^{1+}B^{5+}O_3^{2-}$ and $A^{2+}B^{4+}O_3^{2-}$ compounds have larger $dR_{anti}/dP$ than $A^{3+}B^{3+}O_3^{2-}$ materials. However, these two rules are not the full story, as for example they cannot explain why CaTiO$_3$ has a larger $dR_{anti}/dP$ than CaSnO$_3$ while the tolerance factor of the latter is smaller than that of the former, both belonging to the $A^{2+}B^{4+}O_3^{2-}$ family and having the same A cation. As we will show below, this exotic behavior is related to the presence of fully empty low-lying d-states in some $A^{2+}B^{4+}O_3^{2-}$ materials. In addition, SrTiO$_3$ presents a larger $dR_{anti}/dP$ value than CaTiO$_3$, which is also against the usual trend that $dR_{anti}/dP$ decreases with the tolerance factor for a given series. This is because the magnitude of $dR_{anti}/dP$ is enhanced in systems with small tiltings, as we will discuss later in detail.

Although $dR_{in}/dP$ presents essentially the same trends as $dR_{anti}/dP$, there is a notable difference: the magnitude of $dR_{in}/dP$ (i.e., $|dR_{in}/dP|$) is larger than that of $dR_{anti}/dP$ for systems having large tolerance factor. For example, $|dR_{in}/dP|$ is much larger than $|dR_{anti}/dP|$ in SrTiO$_3$ and LaAlO$_3$. We show below that this is because in-phase tilts are much smaller than antiphase tilts in compounds with large tolerance factor.

It is also important to recall that some perovskite materials might display antiphase tiltings about two or three different pseudo-cubic <001> directions. For instance, LaAlO$_3$ adopts a rhombohedral structure ($R\bar{3}c$ space group) with tiltings about all three pseudo-cubic <001> axes. Our calculations (Fig. S1 of Suppl. Mat.) indicate that the effect of pressure on the antiphase tiltings in $R\bar{3}c$ perovskites is qualitatively similar to the one for rotations about a single <001> axis, depicted in Fig. 1b.

### B. Cases combining anti-phase and in-phase rotations

Interestingly, in-phase and antiphase tiltings are simultaneously present in many perovskites. In fact, such is the case of the most common perovskite structure, the so-called GdFeO$_3$-type. This phase is orthorhombic with the Pbnm space group, and results from the condensation of an antiphase tilt about the [110] pseudo-cubic axis ($R_5^-$ mode) and an in-phase rotation about the [001] pseudo-cubic axis ($M_2^+$ mode) with respect to the ideal cubic perovskite structure. According to Glazer's notation[9], the tilting in the Pbnm structure can therefore be described as $a^-a^-c^+$. In this Pbnm structure, A-site anti-polar displacements ($X_5^-$ mode) are allowed by symmetry, which optimize the A-site cation coordination environment and further stabilize this phase [29-33].

Our extensive DFT calculations on Pbnm compounds reveal that orthorhombic perovskites can adopt rather different pressure behaviors of octahedral tiltings. To demonstrate that, Fig. 2 shows the pressure dependence of the magnitude of the two octahedral tiltings ($R_5^-$ and $M_2^+$ mode) in four selected materials, namely LaFeO$_3$, LuFeO$_3$, CaTiO$_3$ and CaSiO$_3$. The symmetry-mode decomposition is carried out with the ISODISTORT program [34]. Pressure suppresses both in-phase and antiphase tiltings in LaFeO$_3$. In contrast, we find that in the case of CaTiO$_3$ pressure enhances both in-phase and antiphase rotations, in agreement with the experimental result of Ref. [15]. Strikingly, the pressure behavior of the tilts in LuFeO$_3$ is yet different: pressure suppresses the antiphase mode but enhances the in-phase rotation. Note also that other test calculations (not shown here) indicate that Pbnm $A^{3+}B^{3+}O_3^{2-}$ compounds with a small A-site ion (i.e., LuAlO$_3$, TmFeO$_3$, LuCrO$_3$) displays a similar behavior to LuFeO$_3$. Therefore, besides the two well-known behaviors that pressure enhances or suppresses both in-phase and antiphase tiltings, we discover that it can also suppress the antiphase $R_5^-$ mode and enhance the in-phase $M_2^+$ mode in Pbnm compounds with a small tolerance factor.

Figure 2 further indicates that, for CaSiO$_3$, hydrostatic pressure suppresses both $R_5^-$ and $M_2^+$ modes. This is in agreement with the computational work of CaSiO$_3$ [23], but contradicts the general rule proposed by Angel *et al.* [22] which states that the octahedral tiltings in all $A^{2+}B^{4+}O_3^{2-}$ perovskites are enhanced under pressure.

### C. Landau-like description

Let us now introduce an elementary Landau-like potential to describe the energetics of the in-phase and antiphase tilting instabilities, which will be useful for the discussion that follows.

Since octahedral rotations can either be in-phase or anti-phase, and can also be along three different pseudocubic <001> directions, there are six elementary octahedral tilting modes, that are $R_{in}^x, R_{in}^y, R_{in}^z$ (which represent the in-phase tilt, $M_2^+$, about the pseudo-cubic [100], [010] and [001] axis, respectively), and $R_{anti}^x, R_{anti}^y, R_{anti}^z$ (which are associated with the anti-phase tilt, $R_5^-$). Using group theory, we can derive the energy series (up to the fourth order in the tilting amplitudes) for the distorted perovskite structure:

$$E_{tilt} = A_2 \sum_\alpha (R_{in}^\alpha)^2 + A_4 \sum_\alpha (R_{in}^\alpha)^4 + B_2 \sum_\alpha (R_{anti}^\alpha)^2 + B_4 \sum_\alpha (R_{anti}^\alpha)^4 + E_{cross}^4 \qquad \text{Eq. (1)},$$

where $\alpha = \{x, y, z\}$, the coefficients $A_2$ and $A_4$ (respectively, $B_2$ and $B_4$) describe the energy landscape for in-phase (respectively, antiphase) tilts, and $E_{cross}^4$ gathers all the fourth order (including biquadratic) coupling terms between these six tilting modes. The six elementary octahedral tilting modes are adopted as independent variables in order to make this Landau-like potential as general as possible. Our DFT calculations show that all these fourth order coupling terms $E_{cross}^4$ are positive and increase with pressure for all perovskites considered in this work, suggesting that the antiphase and in-phase tilt modes compete with each other and that this competition is enhanced by pressure. Since the pressure behavior of $E_{cross}^4$ are similar to that of the fourth order $A_4$ and $B_4$ terms, we will not discuss them hereafter.

The $A_2$, $A_4$, $B_2$ and $B_4$ coefficients are fitted to DFT results (see Appendix for the computational details). We plot them in Fig. 3, as a function of the pseudo-cubic 5-

atom-cell lattice constant, for the four materials studied in Fig. 2. It can be seen that the fourth order coefficients ($A_4$ and $B_4$) are always positive and increase with pressure (i.e., when decreasing the lattice constant), which is expected since the octahedral tilting will reduce the distance between the next-nearest-neighboring (NNN) oxygen ions, as well as the A-O distance, resulting in a stronger short-range (Pauli) repulsion associated to overlapping electrons of different atoms. The second order coefficients ($A_2$ and $B_2$) are negative, as consistent with instability of the tilts, and display a much richer behavior. They decrease for LaFeO$_3$ and CaSiO$_3$, resulting in weaker tilting instabilities for increasing preasure. The reverse trend occurs in CaTiO$_3$ and LuFeO$_3$. Note that, the tilt magnitude in a $a^0a^0c^-$ (resp. $a^0a^0c^+$) structure is given by $R_{anti}^{min} = \sqrt{\frac{-B_2}{2B_4}}$ (resp. $R_{in}^{min} = \sqrt{\frac{-A_2}{2A_4}}$). Since $|A_2|$ and $|B_2|$ decrease with pressure in LaFeO$_3$ and CaSiO$_3$, and the fourth order coefficients always increase with pressure, it follows that the O$_6$-rotations must decrease under compression, as shown in Fig. 2. In contrast, for CaTiO$_3$ and LuFeO$_3$ there is a competition between the second- and fourth-order terms. In CaTiO$_3$ the tilts increase with pressure because the change of the second-order coefficient is faster than that of the fourth order coefficient. The case of LuFeO$_3$ is more subtle because the behavior of the second- and fourth-order terms indicated in Fig. 3 should result in a decrease of both in-phase and antiphase tilts; this is consistent with the result for single-tilt cases shown in Fig. 1, but contradicts our results for the Pbnm structure in Fig. 2. This apparent contradiction hints at an interaction between in-phase and antiphase tilts in LuFeO$_3$, which, as we will see later, corresponds to the existence of a trilinear coupling involving both octahedral tilts together with an antipolar distortion mode.

### III. DISSCUSSION

Let us now try to better understand the results displayed in Figs. 1 and 2 and unravel the origins of the different pressure behaviors of the oxygen octahedral tiltings. For that, it is important to consider the following aspects.

## A. Origin of the octahedral tilting instabilities

*Dominant interatomic interactions that drive the octahedral tilt instabilities.* It is widely accepted that the instability of octahedral tilting in ABO$_3$ perovskites is due to the tendency to increase the A-O interactions, either covalent or electrostatic [10]. Let us test such a notion against our first-principles calculations.

To do this, we first propose an original way to decompose the second-order Landau coefficients [A$_2$ and B$_2$ in Eq. (1)] into different contributions. Note that these coefficients are directly related to the (harmonic) force constants describing the energy changes for small distortions of the ideal cubic perovskite phase of the material. More specifically, we have

$$E_2 = \frac{1}{2} \sum_{lk\alpha} \sum_{l'k'\beta} \Phi_{\alpha\beta}(lk,l'k') u_\alpha(lk) u_\beta(l'k') \quad \text{Eq. (2)},$$

where $u_\alpha(lk)$ is the atomic displacement along the $\alpha$ direction of the $k$-th atom in the $l$-th unit cell, and $\Phi$ are the harmonic force constants. For the particular case of a O$_6$-tilting distortion, only the oxygen atoms move, and we can write $E_2 = E_2^{self} + E_2^{'O-O}$, where the first term comes from the self-interaction of the oxygen atoms ($lk = l'k'$) and the second one includes all the couplings between couples of different oxygens ($lk \neq l'k'$) in Eq. (2). Now, the self-energy of a particular atom can be actually interpreted as an interaction with the rest of the lattice, by means of the acoustic sum rule $\Phi_{\alpha\beta}(lk,lk) = -\sum_{l'k' \neq lk} \Phi_{\alpha\beta}(lk,l'k')$ where $l'k' \neq lk$ in the sum. Hence, we can use this expression to expand $E_2^{self}$ and, by grouping together the terms involving O—O, O—A, and O—B atomic pairs, we can split the energy as $E_2 = E_2^{A-O} + E_2^{B-O} + E_2^{O-O}$, where it should be noted that $E_2^{O-O}$ contains the interactions in the $E_2^{'O-O}$ energy introduced above, plus additional contributions coming from the self-energy.

In terms of the decomposed energies, one can write the individual contribution to B$_2$ as $B_2^{p-p'} = E_2^{p-p'}/R^2$, where $R$ is the amplitude of the tilting and $p-p'$ represents a particular pair of atom types. Note that negative values of $B_2^{p-p'}$ indicate that the $p-p'$ interaction favors octahedral tilting. The separated contributions to $B_2$ for the four compounds in Fig. 2 are shown in Fig. 4. We can see that there is a significant B-O contribution to $B_2$ besides the expected A-O and O-O contributions. Moreover, $B_2^{A-O}$ increases under compression while $B_2^{B-O}$ decreases and $B_2^{O-O}$ is weakly pressure dependent. Interestingly, both $B_2^{A-O}$ and $B_2^{B-O}$ can change sign as a function of pressure. In particular, for LaFeO$_3$, $B_2^{A-O}$ becomes positive for lattice constants smaller than 3.75 Å, indicating that the A-O interaction is then unfavorable for the occurrence of octahedral tilts. At the same time, in LaFeO$_3$ $B_2^{B-O}$ becomes negative and larger in magnitude than $B_2^{A-O}$, implying that the octahedral tilt instability is now driven by the B-O and O-O interactions (see Fig. 4a). As a result, the B-O contribution becomes more and more important for the condensation of oxygen octahedral tiltings under pressure in LaFeO$_3$. Note that a similar conclusion can be drawn for CaTiO$_3$ and to a lesser extend to LuFeO$_3$, according to Fig. 4.

In order to understand how the B atoms drive the octahedral tilt, we analyze the force constants in more detail. Around each oxygen, there are two nearest-neighboring (NN) B ions and eight NNN B ions ( B' and B" in Fig. 4e, respectively). Without loss of generality, we assume that the central oxygen atom is moving along the *x*-direction when the tilt occurs. Taking LaFeO$_3$ with the lattice constant of 3.8 Å as a typical example, we find that force constant between B' and O along the *x*-direction is $\Phi_2 = -0.78$ eV/Å2, while the coupling between B" and O along the *x*-direction is $\Phi_3 = 1.01$ eV/Å$^2$. For comparison, the force constant between O and its NN A ion along the *x*-direction is $\Phi_1 = 0.16$ eV/Å$^2$. A negative value of $\Phi_2$ results in a positive contribution to $E_2^{B-O}$ and is consistent with the observation that the octahedral tilt makes the NN B-O interaction less favorable. Surprisingly, the force constant between O and its NNN B ion is even larger than that between O and its NN A ion, and of opposite sign; hence, upon condensation of the tilting distortion, the energy reduction associated to the NNN B-O interaction will be larger than the energy penalty coming from the NN A-O interaction. Note that the tilting will bring the O atoms and its NNN B atoms closer, thus optimizing the covalent and/or electrostatic factors contributing to the interaction. When the A-site has a large ionic radius or the B-ion could form a strong covalent bond with oxygen, the NNN B-O interaction is more important than the NN A-O interaction even though the NNN B-O distance is larger than the NN A-O distance. Therefore, we find that the interaction between the NNN B-ion and oxygen ion is another important source of the octahedral tilting instability. Note that this may explain why the octahedral tilting also occurs in materials that, like α-AlF$_3$ [35], do not contain any A-site cations. Interestingly, the effect of the interaction between the NNN B-ion and oxygen ion on octahedral tilting becomes more and more significant under pressure. This also explains our numerical finding that pressure enhances octahedral tiltings in α-AlF$_3$.

*Why does the anti-phase tilting usually have a stronger instability than the in-phase tilting?* Experiments show that antiphase tilts occur more frequently than in-phase tilts in perovskites. For example, SrTiO$_3$ takes the I4/mcm tetragonal structure with a single anti-phase tilt at low temperature. In contrast, to our best knowledge, a perovskite compound never adopts the structure with a single in-phase tilt as the ground state. This is because the instability of the antiphase tilt is stronger than that of the in-phase tilt (that is, $B_2 < A_2$) for a given lattice constant, as shown in Fig. 3. By computing the electrostatic energy with the Ewald method, we find that this is caused by a larger gain in O-O electrostatic energy in the antiphase case. By decomposing the second-order coefficients $A_2$ and $B_2$ into different contributions with the use of second-order force constants, we find that $A_2^{A-O} = B_2^{A-O}$ and $A_2^{B-O} = B_2^{B-O}$, while $A_2^{O-O} > B_2^{O-O}$. Therefore, the O-O interaction favors antiphase tilt over in-phase tilt.

### B. Pressure-dependence of simple oxygen octahedral rotational patterns

*Influence of ionic sizes*: Let us now explain why $dR/dP$ for the compounds in a given series usually decreases with the tolerance factor, as shown in Fig. 1. A small tolerance factor indicates that B and oxygen atoms will be tightly packed while the A atoms will be relatively loose. This suggests that the $BO_6$ octahedron will be less compressible than the $AO_{12}$ polyhedron for perovskites with small tolerance factor. When pressure is applied, the material will thus tend to shorten the A-O bonds while maintaining the distance between B and O atoms, i.e., the octahedral tiltings will tend to increase. If the tolerance factor is large, the opposite applies and pressure suppresses the octahedral tilting. This argument is in agreement with the fact that $B_2$ for $LaFeO_3$ decreases with the lattice constant, while $B_2$ for $LuFeO_3$ increases (see Figs. 3a and b). Figure 4 shows that this is because $B_2^{A-O}$ increases much faster with decreasing lattice constant in $LaFeO_3$ than in $LuFeO_3$, which probably reflects the fact that the electrons of the relatively large $La^{3+}$ cations quickly start to repel the $O^{2-}$ anion under pressure. Thus, this electronic repulsion is the most likely explanation for the observation that $dR/dP$ usually decreases with the tolerance.

*Influence of ionization states:* As we discussed above, pressure usually suppresses the octahedral tilting in $A^{3+}B^{3+}O_3^{2-}$ compounds, while enhancing it in $A^{2+}B^{4+}O_3^{2-}$ materials (see Fig. 1). Such an effect was previously explained in terms of the bond-valence parameters [22]. Here we would like to rather suggest that this effect is due to the dependence of the A-O interaction on the formal charge of the A ion. Generally speaking, the ionic radius of A cations with a high valence is smaller than that of cations with a low valence [27] since more strongly charged cations will tend to move closer to the anion in order to lower the electrostatic energy. For example, the ionic radii of $Rb^{1+}$, $Sr^{2+}$, $Y^{3+}$ are 1.66 Å, 1.32 Å, 1.04 Å, respectively. The distance between oxygen and A ions with a high valence is thus already small at zero pressure, implying that the corresponding A-O bond will be relatively hard to compress. Such a notion is demonstrated in Fig. S3, where $B_2^{A-O}$ is found to increase much faster with decreasing volume in $LaAlO_3$ than in $CaGeO_3$, despite of the fact that they have similar tolerance factors. This explains why, for materials with similar tolerance factors, $\frac{dR}{dP}$ usually decreases as we move from $A^{1+}B^{5+}O_3^{2-}$ to $A^{2+}B^{4+}O_3^{2-}$ and then to $A^{3+}B^{3+}O_3^{2-}$.

*Influence of orbital hybridizations:* Figures 1 and 2 show that $\frac{dR}{dP}$ for CaTiO$_3$, CaZrO$_3$, CaHfO$_3$ and SrTiO$_3$ is much larger than for CaBO$_3$ (B = Si, Ge, Sn or Mn) and SrBO$_3$ (B = Ge or Mn), at variance with the usual trend with respect to the tolerance factor. Interestingly, the key difference between these compounds is that CaTiO$_3$, CaZrO$_3$, CaHfO$_3$ and SrTiO$_3$ have empty low-lying d states, while the others do not. We thus decided to examine whether the empty d states play a role on the effect of pressure on octahedral tiltings.

To isolate the effect of the hybridization between the empty d states and the O-2p states on the octahedral tilting, we employ the orbital selective external potential (OSEP) method [36,37], in which an external field is applied to shift the energy levels of some chosen orbitals. More precisely, we shift the O-2p states to a lower level so that their hybridization with the empty states of the B ion decreases. As can be seen from Fig. 5a, $dR_{anti}/dP$ for CaBO$_3$ (B = Ti, Zr and Hf) becomes smaller for lower-lying O-2p levels. The opposite trend is observed for CaBO$_3$ compounds (B = Si, Ge and Sn). Therefore, it is clear that the hybridization between the empty d states of the B ion and the O-2p states is a key factor leading to larger $dR_{anti}/dP$ values for CaTiO$_3$, CaZrO$_3$ and CaHfO$_3$.

Let us now try to understand why this is the case. For that, we carry out tight-binding (TB) calculations, considering a $R\bar{3}c$ phase of CaTiO$_3$ as a representative model system. In our TB Hamiltonian we consider the Ti-3d orbitals and O-2p orbitals. The hopping integrals $t$ are evaluated with the Slater-Koster scheme [38]. Following Harrison [39], we adopt $t \propto \frac{1}{d^{5/2}}$ to take into account the dependence on the interatomic distance. We compute the energy given by this model (this is simply the band energy $E_b = \sum_{\bm{k}} \sum_{i=1}^{n_{occ}} \varepsilon_{i\bm{k}}$, where $n_{occ}$ is the number of occupied bands and $\varepsilon_{i\bm{k}}$ is the eigenvalue of the *i*-th band at k-point **k** for two cases). In the first case, we only consider the NN Ti-O hopping $t_{NN}$, while in the second case we also include the NNN Ti-O hopping $t_{NNN} = t_{NN}\left(\frac{d_{NN}}{d_{NNN}}\right)^{5/2}$, where $d_{NN}$ and $d_{NNN}$ are the NN Ti-O distance and NNN Ti-O distances, respectively. We thus find that the NNN B-O interaction makes $B_2$ much more negative when the lattice constant is small, i.e., under pressure (see Fig. 5b), which can be explained as follows. When a pressure is applied, the NNN Ti-O interactomic distance is reduced, and the hopping between the O-2p orbitals and Ti-3d orbitals is thus enhanced significantly. Therefore, the hybridization between the Ti-3d and the O-2p orbitals makes the second order coefficients ($A_2$ and $B_2$) more negative, increasing the tendency of octahedral tilting under pressure. Note that when there are too many filled d-electrons, the NNN B-O interaction might increase $B_2$ for decreasing lattice constant (see Fig. 5b for the 4 and 5 d-electrons/B-site cases). This explains why the p-d hybridization plays a less important role on the octahedral tilts in RCrO$_3$ and RMnO$_3$ systems. Hence, this analysis suggests that having empty low-lying d states plays a significant role in the pressure behavior of octahedral tilting in perovskites. Recently, it was pointed out that the tolerance factor alone does not determine the temperature at which the cubic phase is stabilized and that the electronic configuration of the B-site cation appears to also be of significance [40]. This is in agreement with our present finding that the d-orbital occupation is also relevant to the octahedral tilting. By considering the NN B-O interaction, Woodward proposed that the B-O π-bonding favors the cubic perovskite ABO$_3$ structure if the π* t$_{2g}$ d-band is filled and less than half-filled [10]. In contrast, here we consider the role of the NNN B-O interaction.

*Influence of magnitude of the octahedral tilting:* Figure 1 shows that SrTiO$_3$ has a larger $dR/dP$ than CaTiO$_3$, and that $|dR_{in}/dP|$ is much larger than $|dR_{anti}/dP|$ in SrTiO$_3$, LaAlO$_3$, CaSiO$_3$ and SrMnO$_3$. To be more specific, $dR_{in}/dP$ is more negative than $dR_{anti}/dP$ in LaAlO$_3$, CaSiO$_3$ and SrMnO$_3$, while $dR_{in}/dP > dR_{anti}/dP > 0$ in

SrTiO$_3$. These unusual behaviors can not be explained by the three effects we just discussed above. An indication appears to be that this phenomenon happens when the tilt at zero pressure is small. For simplicity, let us consider the in-phase tilting case. Since the magnitude of tilt at zero pressure can be computed as $R_{in} = \sqrt{\frac{-A_2}{2A_4}}$, we have $dR_{in}/dP = \frac{1}{2R_{in}} d(\frac{-A_2}{2A_4})/dP$. This suggests that the magnitude of $dR_{in}/dP$ is inversely proportional to the magnitude of the tilts. In fact, the in-phase tilts in all these four compounds (SrTiO$_3$, LaAlO$_3$, CaSiO$_3$ and SrMnO$_3$) are numerically found to be very small. Our DFT calculations also predict that CaTiO$_3$ has an even slightly larger $\left| d(\frac{-A_2}{2A_4})/dP \right|$ than SrTiO$_3$ (not shown here), which further suggests that the reason why SrTiO$_3$ has a larger $\left| dR_{in}/dP \right|$ than CaTiO$_3$ is that SrTiO$_3$ has a much smaller in-phase tilt than CaTiO$_3$.

### C. Cases combining in-phase and antiphase rotations

So far we have discussed the behavior of simple structures in which only one type of tilting pattern, in-phase or antiphase, exists. The observed trends should be applicable to cases combining in-phase and antiphase rotations, as those shown in Fig. 2. However, there is another effect displayed in Fig. 2 that cannot be explained by the above considerations, that is, why pressure suppresses the antiphase $R_5^-$ mode but simultaneously enhances the in-phase $M_2^+$ mode in Pbnm LuFeO$_3$. First, it is important to realize that, according to our numerical results, the strain degree of freedom is not the key to this behavior, as evidenced in Fig. S4 of Suppl. Mat. Therefore, we hereafter focus on the cubic cell for simplicity, and recall that the Pbnm state also displays antipolar motions of the A cations. Interestingly, we numerically find that if we suppress such antipolar motions, by fixing the A-ion to their ideal high-symmetry positions, pressure suppresses both the antiphase mode and in-phase rotations in Pbnm LuFeO$_3$, in agreement with the behavior of the tilting patterns considered individually that we found for this compound (see Figs. 1a and 1b). Therefore, a coupling between the A-site related antipolar mode (of $X_5^-$ symmetry) with the rotational modes ($R_5^-$ and $M_2^+$) should be responsible for the exotic behavior of LuFeO$_3$ under pressure. Interestingly, it has been recently shown that there is a specific *trilinear* coupling between the three modes existing in any Pbnm perovskite [29]. Incorporating such coupling in a simple model gives:

$$E = E_{tilt} + E(X) + E_c,$$

with $E_{tilt} = E(R) + E(M) + E(R,M)$ and $E_c = E(X,R) + E(X,M) + \lambda XRM$,

where $R$ and $M$ represent the amplitude of the tilting patterns $a^-a^-c^0$ ($R_5^-$) and $a^0a^0c^+$ ($M_2^+$), respectively; $X$ denotes the amplitude of the $X_5^-$ mode; and we also have individual mode energies $E(M) = A_2^M M^2 + A_4^M M^4$, $E(R) = B_2^R R^2 + B_4^R R^4$ and $E(X) = C_2^X X^2 + C_4^X X^4$; finally, $\lambda$ denotes the strength of the trilinear coupling and note that, for simplicity, we do not consider strains in the model. The fourth order repulsion terms $E(P,Q)$ take the form of $P^2Q^2$ ($P,Q = X, R$ or $M$, and $P \neq Q$). Note that $B_2^R$ and $B_4^R$ are related to, but different from, $B_2$ and $B_4$ in Eq. (1), since the $a^-a^-c^0$-like distortion is an antiphase octahedral rotation about *both* the pseudo-cubic *x* and *y* axes.

By fitting this model energy to DFT results, we obtain the corresponding parameters at different pressures. Note that for these DFT simulations we consider structures in which the cell is forced to be cubic, but the atoms move from their high-symmetry positions as in a regular Pbnm phase. At each pressure, the lattice constant of the cubic cell is chosen so that the cell volume is the one obtained for the actual Pbnm structure of LuFeO$_3$ with an orthorhombic cell. Some of these fitted parameters are shown in Fig. 6a. The fourth order coefficients $A_4^M$ and $B_4^R$ are positive and increase with pressure. It is important to note that $B_4^R$ grows faster than $A_4^M$, suggesting that the antiphase tilt becomes less favorable than the in-phase rotations under pressure. We also find (not shown here) that (i) as expected, $A_2^M$ and $B_2^R$ decrease with pressure, in agreement with our above results for the second order coefficients $A_2$ and $B_2$ in Fig. 3; and (ii) $C_2^X$ is negative, indicating that the $X_5^-$ mode itself is an instability of the cubic phase of LuFeO$_3$. This is different from the usual case (e.g., LaGaO$_3$) where the $X_5^-$ mode itself is stable and its occurrence in the Pbnm structure is induced by the trilinear coupling [29,33]. The small size of Lu$^{3+}$ (and the small tolerance factor of LuFeO$_3$) are surely responsible for this behavior. Interestingly, we also find that the magnitude of the trilinear coupling $|\lambda|$ increases rapidly with pressure (see Fig. 6a). Furthermore, in Fig. 6b we report the amplitude of the $R_5^-$ and $M_2^+$ modes as obtained from the DFT calculations and reproduced by our simple model. While the qualitative agreement is good, there are quantitative discrepancies that are probably due to the 4$^{th}$-order truncation of our model potential.

Let us now discuss the origin of the differentiated behavior of in-phase and antiphase tiltings under pressure in LuFeO$_3$. We numerically found that, when removing the pressure dependence of the trilinear coupling (i.e. making $\lambda$ constant equal to its value at zero pressure), both the in-plane and antiphase tilts are suppressed by pressure. Further, the reason why the pressure enhances the in-phase $M_2^+$ mode, but suppresses the antiphase $R_5^-$ mode, is that the $B_4^R$ coupling increases faster with pressure than $A_4^M$. Indeed, if we make the pressure dependence of $B_4^R$ identical to that of $A_4^M$, our model predicts that pressure would then enhance the $R_5^-$ mode and suppress the $M_2^+$ distortion. Furthermore, increasing the pressure dependence of $A_4^M$, to make it equal to that of $B_4^R$, leads to a suppression of both the $R_5^-$ and $M_2^+$ distortions.

Hence, we find that the peculiar behavior of LuFeO$_3$ relies on a complex interplay among the anharmonic couplings $A_4^M$, $B_4^R$ and $\lambda$ and their pressure dependence. In particular, the enhancement of the tri-linear coupling under pressure, and faster hardening of the antiphase distortions, are responsible for the observed behavior of this compound.

### D. Rules for the pressure effect on octahedral tilting

From our above discussion, we are now in a position to propose the following rules governing the effect of pressure on octahedral tilting in perovskites. (I) The derivative of the tilting angles with respect to pressure ($\frac{dR}{dP}$) decreases when increasing tolerance factor. (II) For materials with similar tolerance factors, $\frac{dR}{dP}$ increases as we move from $A^{3+}B^{3+}O_3^{2-}$, to $A^{2+}B^{4+}O_3^{2-}$, and finally to $A^{1+}B^{5+}O_3^{2-}$ compounds. (III) Perovskites in which the B-site transition metal presents low-lying empty d-states display relatively large values of $\frac{dR}{dP}$, as compared to similar compounds (similar tolerance factors, same nominal ionization states) in which this condition is not fulfilled. (IV) Materials with small tilting instabilities tend to display larger $\left|\frac{dR}{dP}\right|$ values than similar compounds (similar tolerance factors, same nominal ionization states) in which the $O_6$ rotations are large.

Note that Rules I and II were implied in the literature, while Rules III and IV are, to the best of our knowledge, proposed here for the first time. Furthermore, the physical origins of all four rules are revisited and/or unveiled in the present work. For instance, Rule I follows from the fact that the A-site driven octahedral tilting instabilities decrease quickly under pressure if the A-cation is large. Rule II originates from the fact that A-cations in a high ionization state have smaller radii, and are harder to compress, than A-ions in a low ionization state. We find that Rule III applies because NNN B-O covalent interactions are enhanced when octahedral tilt increases under pressure. Finally, Rule IV follows from the fact that the magnitude of $dR/dP$ is roughly inversely proportional to the magnitude of tilt.

In addition, we observe that pressure can also enhance the in-phase octahedral tilting, but suppress the antiphase octahedral tilting, in orthorhombic perovskites having a small tolerance factor. Such an effect strongly relies on the trilinear energy coupling among the in-phase tilt, the antiphase tilt, and an antipolar distortion of the A cations.

### E. Relationship with phonon spectrum

Experimentally, the dependency of the frequency of some phonon modes on pressure has often been employed to deduce the effect of pressure on the octahedral tilts. As shown in Figs. S4 and S5 of the Suppl. Mat., the softening of the low-frequency tilt-related phonon mode always indicates that pressure suppresses the tilt. However, the hardening of the tilt-related phonon modes is not necessarily accompanied by an enhancement of tilt, as evidenced in the Supplemental Material.

### F. Further Applicability of the formulated rules

In this work we focus on perovskite oxides. However, the proposed basic rules should be applicable to other perovskite systems, such as the compounds of the $ABF_3$ family. For example, we performed DFT calculations on $NaMgF_3$ and found that the octahedral tilting increases with pressure, in agreement with the experimental result of Ref. [41]. Since the tolerance factor is small (0.943) and the B-site ion ($Mg^{2+}$) has a higher valence than the A-site ion ($Na^{1+}$), this result is in agreement with Rules I and II.

Note that we did not consider the effect of ferroelectric distortions on octahedral tiltings. Since most perovskites are not ferroelectric and the coexistence of ferroelectricity and octahedral tilts is quite rare, our rules should be applicable to a large number of these materials. We should however note that ferroelectricity and octahedral tilts coexist in some systems (e.g., R3c $BiFeO_3$), where the pressure effect on the tilts is left for future investigations. In addition, our rules may not be applicable to orbital-ordered systems, where the coupling between Jahn-Teller distortion and octahedral tilts is also expected to play a role.

As a demonstration of the further applicability of our formulated rules, we will now examine how the so-called hybrid improper ferroelectricity can be affected by pressure. For that, it is important to recall that a trilinear coupling between the two types of octahedral rotation and a polar mode was recently suggested to give rise to this "hybrid improper ferroelectricity" [30,32,33,42] in ordered perovskites [30,33] and Ruddlesden–Popper compounds [32]. Since the polar mode is induced by two rotational modes, it is expected from our present work that the pressure can tune the hybrid improper ferroelectricity indirectly, by controlling the amplitude of the octahedral rotations. As shown in the Fig. S7 of Suppl. Mat., we demonstrate this point in 1:1 superlattices made of $CaSiO_3$-$MgSiO_3$ and $LaGaO_3$-$YGaO_3$. In fact, our DFT calculations show that pressure enhances the polarization in the hybrid improper ferroelectricity in $CaSiO_3$-$MgSiO_3$, but suppresses it in $LaGaO_3$-$YGaO_3$, which can be easily understood by recalling the behavior of the octahedral rotations of the parent compounds (see Fig. S7 of Suppl. Mat.).

Note that the Suppl. Mat. also provides the (subtle) relation between the effect of pressure on octahedral tilting and how strain reacts to this pressure in perovskites, which should be of benefit to experimentalists using X-ray techniques to (indirectly) probe the role of pressure on tiltings.

## IV. SUMMARY

To summarize, we have comprehensively investigated how pressure affects octahedral rotations in perovskite oxides. Our work has allowed us to confirm and explain some of the existing empirical rules proposed to govern these behaviors, as well as to reveal and understand additional trends that, as far as we know, had never been reported before. Thus, our work provides a detailed guide to understand (and predict) the structural response of the vast majority of perovskite oxides under pressure, which should be especially useful given the importance of these effects and the difficulties involved in their experimental characterization. We have also briefly illustrated the implications of our results and conclusions in what regards other materials' families (e.g., fluorides with the perovskite structure) and materials-design problems (e.g., to tune the so-called hybrid improper ferroelectricity). It is expected that the biaxial strain can also affect the octahedral tiltings, which we will leave for a future study.

**APPENDIX: COMPUTATIONAL DETAILS**

Our total energy calculations are based on the density functional theory within the generalized gradient approximation [43] on the basis of the projector augmented wave method [44,45] encoded in the Vienna *ab-initio* simulation package [46,47]. The plane-wave cutoff energy is set to 500 eV. For REFeO$_3$, the Hubbard on-site repulsion [48] is added for Fe 3d orbitals. Following the previous DFT+U studies on similar systems [49-51], the on-site repulsion U and exchange parameter J for Fe are set to 5 and 1 eV, respectively. In the orbital selective external potential (OSEP) approach [36,37], we add an extra potential $H_{add} = |inlm\sigma\rangle\langle inlm\sigma| V_{ext}$ to the original Kohn-Sham Hamiltonian, where $V_{ext}$ is the applied energy shift, *i* denotes the atomic site, and *n, l, m, σ* are the main quantum number, orbital quantum number, magnetic quantum number, spin index, respectively.

The model parameters of the Landau potential are estimated by fitting to the DFT results. We first obtain the parameters for each single mode by performing a series of DFT calculations with different amplitudes of the mode. We then obtain the coupling between two modes by using the DFT total energy of the states with two condensed modes and the previously obtained parameters for the single mode. In the tri-linear coupling case, we finally extract the coupling parameter $\lambda$ using the DFT total energy of the state with the coexistence of three modes and the already obtained parameters.

**ACKNOWLEDGEMENT**

Work at Fudan was supported by NSFC (11374056), the Special Funds for Major State Basic Research (2015CB921700), Program for Professor of Special Appointment (Eastern Scholar), Qing Nian Ba Jian Program, and Fok Ying Tung Education Foundation. Work at Arkansas is supported by the Office of Basic Energy Sciences, under contract ER-46612 and the DARPA grant HR0011-15-2-0038 (MATRIX program). We also acknowledge the FNR Luxembourg Grants P12/4853155 (M.G, J.I. and J.K.) and INTER/MOBILITY/15/9890527 GREENOX (L.B. and J.I.). H. X. thanks Prof. C. G. Duan for sharing the OSEP code and Dr. Ke Xu for a critical reading of the manuscript.

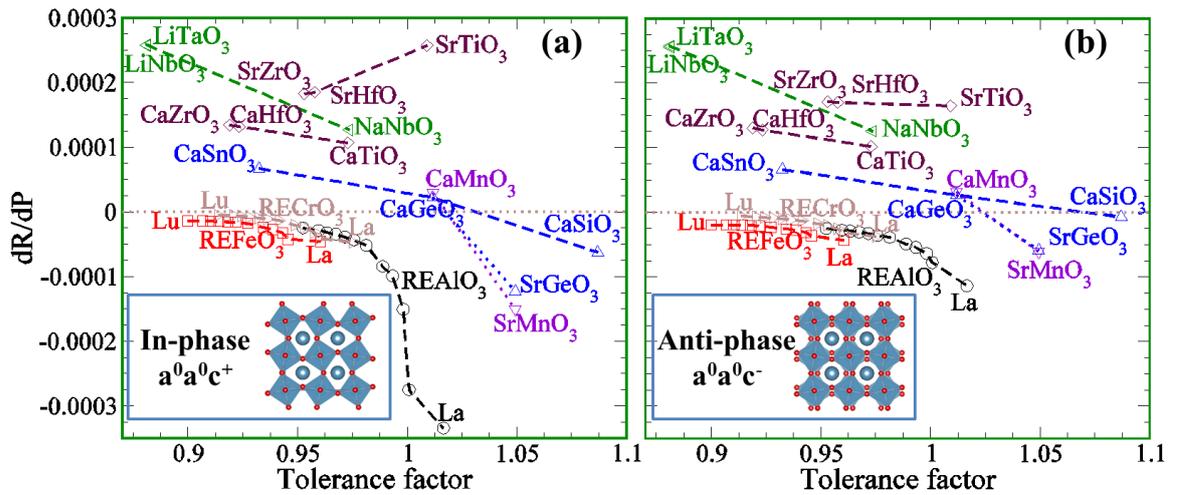

FIG. 1. Derivative of (a) in-phase and (b) anti-phase rotation with respect to pressure as a function of the tolerance factor. The results of several series (ReFeO$_3$, ReAlO$_3$, ReCrO$_3$, CaBO$_3$, SrBO$_3$, LiBO$_3$, and NaNbO$_3$) are shown.

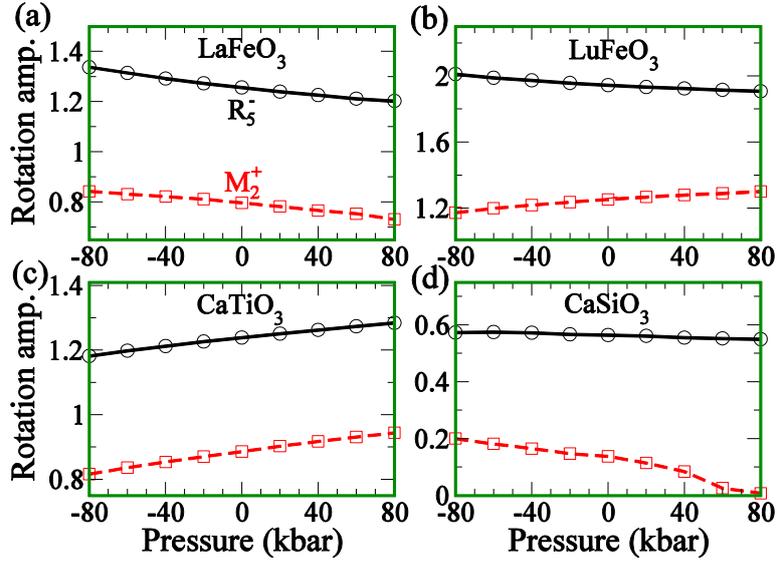

FIG. 2. Pressure effect on octahedral tiltings in Pbnm (a) LaFeO$_3$, (b) LuFeO$_3$, (c) CaTiO$_3$, and (d) CaSiO$_3$. The behaviors of both $R_5^-$ mode (anti-phase tilt about the [110] axis) and $M_2^+$ mode (in-phase tilt about the [001] axis) are shown.

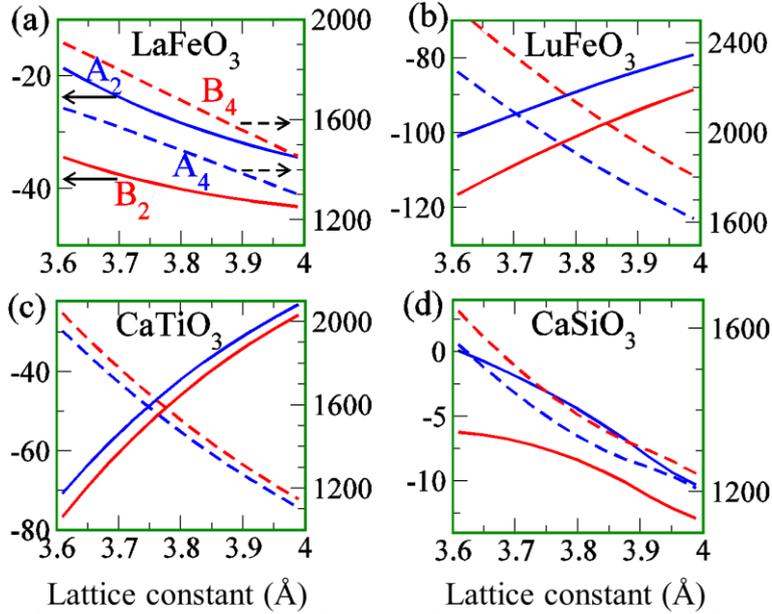

FIG. 3. Dependence of the coefficients ($A_2$, $B_2$, $A_4$, $B_4$) of the energy model [Eq. (1)]

on the lattice constant. The coefficients A$_2$, A$_4$, B$_2$, B$_4$ are in unit of eV/formula-unit. Four cases [(a) LaFeO$_3$, (b) LuFeO$_3$, (c) CaTiO$_3$, and (d) CaSiO$_3$] are shown.

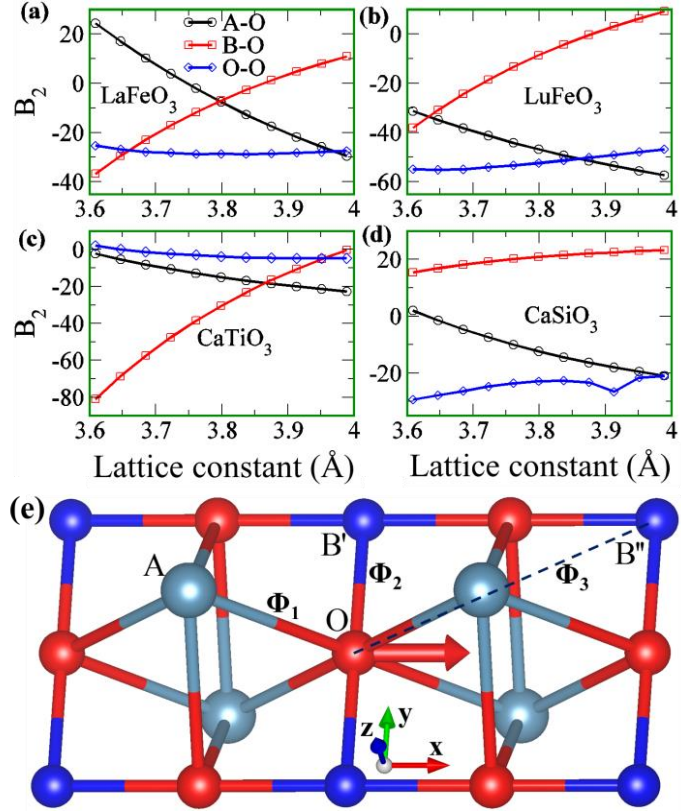

FIG. 4. Contributions ($B_2^{A-O}$, $B_2^{B-O}$, and $B_2^{O-O}$) from different interactions to B$_2$ of the energy model [Eq. (1)] as a function of lattice constant. The coefficients B$_2$ are in unit of eV/formula-unit. Four cases [(a) LaFeO$_3$, (b) LuFeO$_3$, (c) CaTiO$_3$, and (d) CaSiO$_3$] are shown. The local environment for an oxygen moving along the $x$-direction (due to a octahedral rotation about $z$) is shown in Panel (e). $\Phi_1$, $\Phi_2$, and $\Phi_3$ represent the A-O, B'-O, and B''-O force constants, respectively.

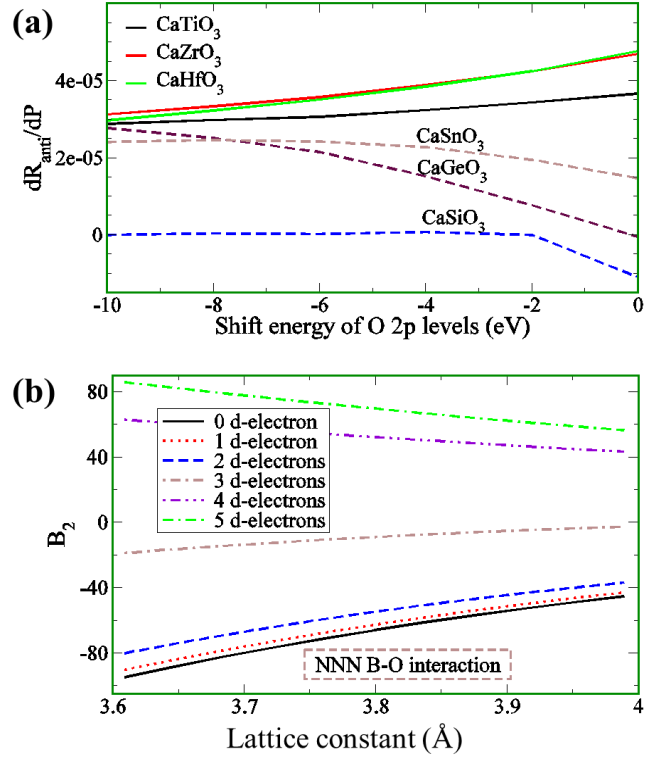

FIG. 5. Origin of the pressure enhancement of octahedral rotations in $ABO_3$ compounds with low-lying d states of B-site. (a) Derivative of anti-phase rotation with respect to pressure as a function of the shifted energy of oxygen 2p states. When the shift energy is more than -2 eV, $CaSiO_3$ becomes cubic with $R_{anti}=0$ and $\frac{dR_{anti}}{dP}=0$. (b) Contribution from the NNN B-O hybridization to $B_2$ of the energy model [Eq. (1)] from the TB calculation. One can see that the NNN B-O interaction makes $B_2$ much more negative when the lattice constant decreases in systems with low occupation of the $d$ orbitals.

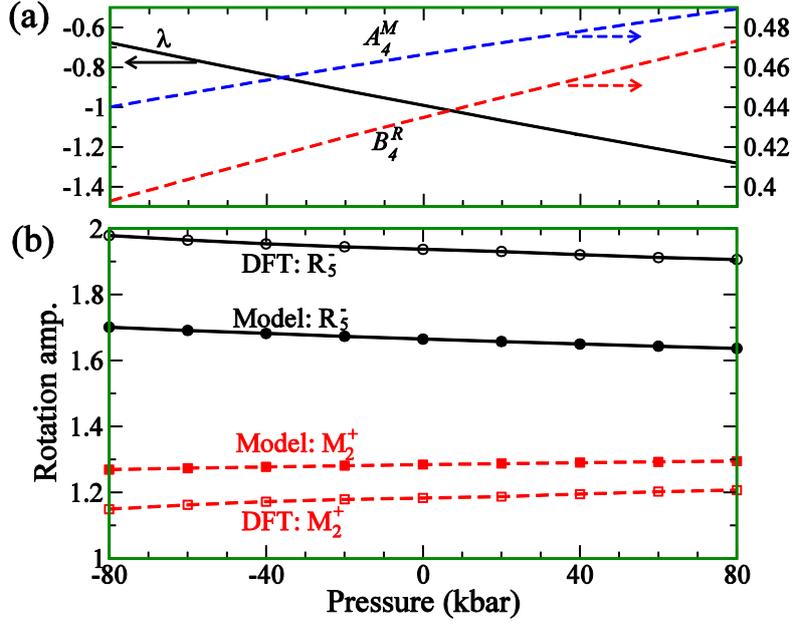

FIG. 6. Explanation for the pressure behavior of octahedral tilt in Pbnm LuFeO$_3$. (a) Parameters ($\lambda$ denotes the trilinear coupling strength, $A_4^M$ and $B_4^R$ are fourth order coefficients of in-phase rotation and anti-phase rotation, respectively) of the energy simple model incorporating the tri-linear coupling, as a function of pressure. (b) Amplitude of the in-phase and antiphase rotations as a function of pressure from the DFT calculation and the energy model. Since the strain degree of freedom is not crucial for the pressure behavior of octahedral tilt (see Fig. S4 of Suppl. Mat.), the cell is kept cubic here.